\documentclass[12pt,a4paper]{article}

\usepackage[utf8]{inputenc}
\usepackage[T1]{fontenc}
\usepackage{lmodern}
\usepackage{amsmath,amssymb}
\usepackage{graphicx}
\usepackage{geometry}
\usepackage{setspace}
\usepackage{caption}
\usepackage{hyperref}
\usepackage{subcaption}
\usepackage{float}
\usepackage[numbers,sort&compress]{natbib}
\usepackage{amsmath,amssymb}
\usepackage{algorithm}
\usepackage{algpseudocode}
\usepackage{booktabs}

\hypersetup{
    colorlinks=true,
    linkcolor=blue,
    citecolor=blue,
    urlcolor=blue
}

\title{A Quantum Genetic Algorithm for Optimization of Benchmark Test Functions using Entangled Qutrits}

\author{
Muhammad Farooq\textsuperscript{1,2,*},
Mehrosh Talat\textsuperscript{2}
\\[8pt]
\small \textsuperscript{1}Normandie Universit\'e,
ENSICAEN, UNICAEN, CNRS, CRISMAT,
14000 Caen, France
\\[5pt]
\small 
\textsuperscript{2}Department of Physics, Air University, 44000 Islamabad, Pakistan
\\[5pt]
\small 
\textsuperscript{*}Corresponding author:
\texttt{muhammad.farooq@ensicaen.fr}
}

\begin{document}
\maketitle
\begin{abstract}
Typically, quantum genetic algorithms are based on qubits, a two-level quantum system. This work investigates the optimization of Rastrigin, Rosen-Brock and Table Holder Function in comparison with Classical Genetic Algorithm (CGA), using two quantum approaches: Quantum Genetic Algorithm (QGA), and Higher-Order Quantum Genetic Algorithm (HO-QGA) based on qutrits, a three-level system. HO-QGA uses higher-dimensional quantum representations, such as entangled qutrits, to test its capability of enhanced optimization. We show that the QGA convergence rate is higher than the CGA, and the solution quality is considerably improved. Based on this study, a three-state quantum object novelly provides more capabilities than qubits in terms of computational and information resources.
\end{abstract}
\tableofcontents
\newpage

\section{Introduction}
Originally developed by John Holland \cite{Holland1973} in the 1970s, classical genetic algorithms (based on the classical bits) are a class of optimization algorithms influenced by the principles of natural selection and evolution. CGAs are helpful when there is a large and complex solution space.

QGA is a quantum-inspired genetic algorithm which uses quantum parallelism and probabilistic properties to solve optimization problems effectively. Genetic and evolutionary algorithms have been extensively investigated as population-based approaches to numerical optimization \cite{Holland1975,DeJong1975,Goldberg1989,Whitley1994,Back1996,EibenSmith2003}. Subsequent studies
have examined convergence, elitism, search diversity,
and algorithm performance across challenging optimization landscapes \cite{Wolpert1997,StornPrice1997,Yao1999,HansenOstermeier2001,Deb2002,Jamil2013}.
Quantum-inspired genetic algorithms introduced probabilistic representations and quantum-inspired evolutionary operators to extend conventional search methods
\cite{Narayanan1996,Han2000,Han2001,Han2002,Han2004}.
More recently, multilevel and qutrit-based quantum
genetic algorithms have been investigated for numerical optimization \cite{Tkachuk2018a,Tkachuk2018b,Hakemi2024}, motivating comparative studies of classical GA, QGA, and higher-order QGA on benchmark functions such as
Rosenbrock, Rastrigin, and Table Holder. Higher-order quantum-inspired representations and their optimization properties have also been examined
\cite{Zhang2011,Nowotniak2014}.

Quantum computations use qubit, a two level system that serves as the basic building block of quantum information:

\begin{equation}
|\psi\rangle = a|0\rangle + b|1\rangle
\end{equation}

\noindent with $a$ and $b$ the probability amplitudes which satisfies the normalization condition:

\begin{equation}
a^2 + b^2 = 1
\end{equation}

\noindent The Qutrit is the unit of quantum for three level system with the state vector:

\begin{equation}
|\phi\rangle = \alpha|0\rangle + \beta|1\rangle + \gamma|2\rangle
\end{equation}

\noindent with the normalization condition:

\begin{equation}
\alpha^2 + \beta^2 + \gamma^2 = 1
\end{equation}

A bipartite state refers to a quantum state that describes a system composed of two subsystems, often denoted as $A$ and $B$. The overall state of the combined system can be represented as a vector in the tensor product of the Hilbert spaces corresponding to the two subsystems, $H_{A} \otimes H_{B}$.

The concept of quantum entanglement was first introduced by Erwin Schrodinger in 1935 \cite{Schrodinger1935a,Schrodinger1935b}. Quantum entanglement refers to a special type of correlation between quantum particles, where the state of one particle is intrinsically connected to the state of another, regardless of the distance separating them. Quantum state is considered entangled if it cannot be expressed as a simple product of the states of the individual subsystems. For the bipartite state, the tensor self-product of the above state is given by:

\begin{equation}
\begin{aligned}
|\phi\rangle \otimes |\phi\rangle
={}& \alpha^2|00\rangle + \alpha\beta|01\rangle + \alpha c|02\rangle\\
&+ \beta\alpha|10\rangle + \beta^2|11\rangle + \beta c|12\rangle\\
&+ c\alpha|20\rangle + c\beta|21\rangle + c^2|22\rangle.
\end{aligned}
\end{equation}

\section{Model and Method}

In this work, three different evolutionary techniques: GA (Algorithm \ref{alg:GA}), QGA (Algorithm \ref{alg:QGA}) and HO-QGA (Algorithm \ref{alg:HOQGA}) are used to optimize three benchmark functions: Rosenbrock, Rastrigin and Holder-table, comparing their convergence rate and accuracy.

\begin{algorithm}[htbp]
\caption{Optimization using CGA}
\label{alg:GA}
\begin{algorithmic}[1]
\State Define benchmark function $f(x,y)$ and search bounds
\State Initialize population $P=10$, generations $G=50$, mutation rate $p_m=0.1$
\State Generate random population of $(x,y)$ solutions
\For{$g=1,\ldots,G$}
    \State Evaluate fitness $f(x,y)$ for all individuals
    \State Select four parents with the lowest fitness
    \State Generate offspring: $\mathbf{c}=(\mathbf{p}_1+\mathbf{p}_2)/2$
    \State Apply mutation with probability $p_m=0.1$
    \State Retain the best individual (elitism)
    \State Update population and record best fitness
\EndFor
\State \Return Best solution and fitness history
\end{algorithmic}
\end{algorithm}

\begin{algorithm}[htbp]
\caption{Optimization using QGA}
\label{alg:QGA}
\begin{algorithmic}[1]
\State Define $f(x,y)$ and divide search range into 10 sections, each with 3 subsections
\State Set population $P=10$, generations $G=50$, mutation rate $p_m=0.1$
\State Initialize normalized qutrit states $|\psi\rangle=(|0\rangle+|1\rangle+|2\rangle)/\sqrt{3}$
\For{$g=1,\ldots,G$}
    \State Generate random number $r\in[0,1)$
    \State Select qutrit outcome using cumulative weights $w_j=|\alpha_j|^2$
    \State Decode selected subsections to obtain $(x,y)$
    \State Evaluate fitness $f(x,y)$
    \State Select the best $50\%$ of the population
    \State Apply crossover, mutation and normalization
    \State Update population and record best fitness
\EndFor
\State \Return Best solution and fitness history
\end{algorithmic}
\end{algorithm}

\begin{algorithm}[htbp]
\caption{Optimization using HO-QGA}
\label{alg:HOQGA}
\begin{algorithmic}[1]
\State Define $f(x,y)$ and construct 10 sections with 9 subsections each
\State Set population $P=10$, generations $G=50$, mutation rate $p_m=0.1$
\State Generate random nine-component vectors $\mathbf{c}\in[0,1)^9$
\State Normalize each state: $\mathbf{c}\gets\mathbf{c}/\|\mathbf{c}\|$
\For{$g=1,\ldots,G$}
    \State Calculate probabilities $w_j=|c_j|^2$, $j=0,\ldots,8$
    \State Select outcomes using cumulative probabilities
    \State Decode selected subsections into $(x,y)$
    \State Evaluate fitness $f(x,y)$
    \State Select the best $50\%$ of the population
    \State Apply crossover and Gaussian mutation ($p_m=0.1$)
    \State Normalize offspring and update population
    \State Record best fitness for convergence analysis
\EndFor
\State \Return Best solution and fitness history
\end{algorithmic}
\end{algorithm}

\noindent The minima of Rosenbrock function lies under a narrow valley whereas rastrigin function have multiple local minimas and one gobal minima but table holder function has 4 global minima. The mathematical forms of these benchmark functions are given by:

\begin{equation}
 f_{\mathrm{Ros}}(x,y)=(1-x)^2+100(y-x^2)^2,
 \label{eq:rosenbrock}
\end{equation}
\begin{equation}
 f_{\mathrm{Ras}}(x,y)=20+x^2+y^2
 -10\cos(2\pi x)-10\cos(2\pi y),
 \label{eq:rastrigin}
\end{equation}
\begin{equation}
 f_{\mathrm{Hol}}(x,y)=-\left|\sin x\,\cos y\,
 \exp\!\left(\left|1-\frac{\sqrt{x^2+y^2}}{\pi}\right|\right)\right|.
 \label{eq:holder}
\end{equation}

\noindent The Rosenbrock function has a unique global minimum at
$(x_1,x_2)=(1,1)$, where $f_{\min}=0$, within the search domain
$-2.048\leq x_i\leq 2.048$. Similarly, the Rastrigin function
has a global minimum at $(x_1,x_2)=(0,0)$, where $f_{\min}=0$,
within the domain $-5.12\leq x_i\leq 5.12$.
The Holder Table function possesses four global minima at
$(x_1,x_2)=(\pm 8.05502,\pm 9.66459)$, where
$f_{\min}\simeq -19.2085$, within the search domain
$-10\leq x_i\leq 10$. Here, $i=1,2$ represents the two
optimization variables.

\section{Results and Discussion}

The performance of GA, QGA, and HO-QGA is compared using benchmark functions in Eqs.~(\ref{eq:rosenbrock}), (\ref{eq:holder}), and (\ref{eq:rastrigin}).
The results highlight convergence behavior and optimization
accuracy. Tables.~(\ref{tab:GA_results}), (\ref{tab:QGA_results}), and (\ref{tab:HOQGA_results}) show the optimum value $f_{opt}$, best fitness value $f_{best}$ achieved in this work at the generation $g_{best}$ and the uncertainty $|\Delta f|$ for each benchmark function. 

\begin{figure}[H]
\centering
\includegraphics[width=0.73\textwidth]{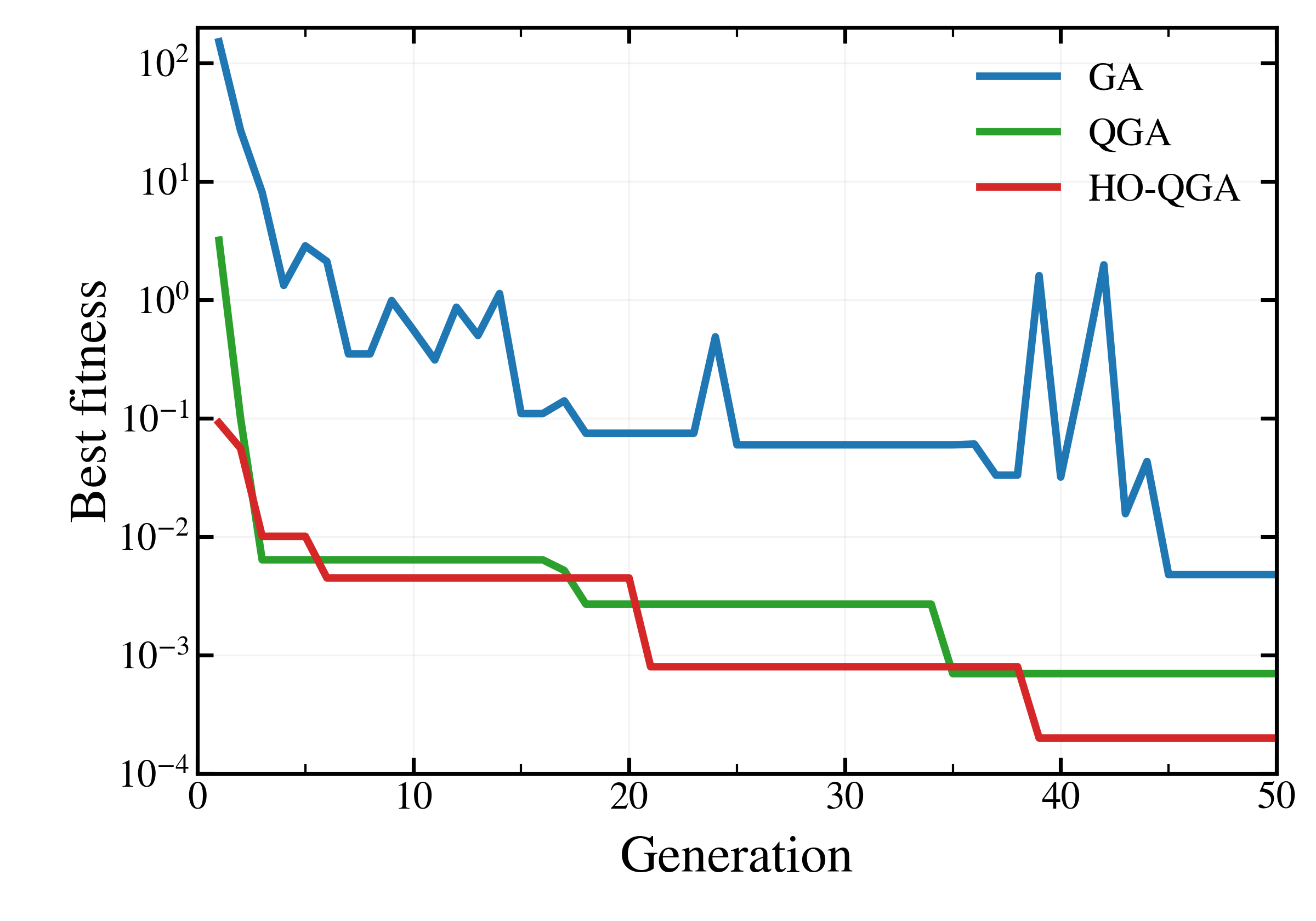}
\caption{Optimization performance of Rosenbrock function using Classical genetic algorithm (GA), quantum genetic algorithm (QGA), and higher-order quantum genetic algorithm (HO-QGA)}
\label{fig:rosenbrock}
\end{figure}

Fig.~\ref{fig:rosenbrock} shows the optimization performance of Rosenbrock function (Eq.~\ref{eq:rosenbrock}), which have a global minimum at $f_{\min}=0$. The initial fitness value of classical GA is $151.9841$ and reaches $0.0048$ at generation $50$ (Table.~\ref{tab:GA_results}). Comparatively, QGA reaches $0.0007$ at generation $35$ (Table.~\ref{tab:QGA_results}) and HO-QGA reaches $0.0002$ at generation $39$ (Table.~\ref{tab:HOQGA_results}).


\begin{table}[htbp]
\centering
\caption{Optimization results using Classical Genetic Algorithm (GA).}
\label{tab:GA_results}
\small
\begin{tabular}{lcccc}
\toprule
Function & $f_{\mathrm{opt}}$ & $f_{\mathrm{best}}$
& $|\Delta f|$ & $g_{\mathrm{best}}$ \\
\midrule
Rosenbrock  & $0$        & $0.0048$   & $0.0048$ & $45$ \\
Rastrigin   & $0$        & $0.0076$   & $0.0076$ & $11$ \\
Holder Table & $-19.2085$ & $-16.8551$ & $2.3534$ & $18$ \\
\bottomrule
\end{tabular}
\end{table}


\begin{table}[htbp]
\centering
\caption{Optimization results using Quantum Genetic Algorithm (QGA).}
\label{tab:QGA_results}
\small
\begin{tabular}{lcccc}
\toprule
Function & $f_{\mathrm{opt}}$ & $f_{\mathrm{best}}$
& $|\Delta f|$ & $g_{\mathrm{best}}$ \\
\midrule
Rosenbrock  & $0$        & $0.0007$   & $0.0007$ & $35$ \\
Rastrigin   & $0$        & $0.0031$   & $0.0031$ & $26$ \\
Holder Table & $-19.2085$ & $-19.2080$ & $0.0005$ & $30$ \\
\bottomrule
\end{tabular}
\end{table}


\begin{table}[htbp]
\centering
\caption{Optimization results using Higher-Order Quantum
Genetic Algorithm (HO-QGA).}
\label{tab:HOQGA_results}
\small
\begin{tabular}{lcccc}
\toprule
Function & $f_{\mathrm{opt}}$ & $f_{\mathrm{best}}$
& $|\Delta f|$ & $g_{\mathrm{best}}$ \\
\midrule
Rosenbrock  & $0$        & $0.0002$   & $0.0002$ & $39$ \\
Rastrigin   & $0$        & $0.0007$   & $0.0007$ & $43$ \\
Holder Table & $-19.2085$ & $-19.2085$ & $0.0000$ & $40$ \\
\bottomrule
\end{tabular}
\end{table}

\begin{figure}[H]
\centering
\includegraphics[width=0.73\textwidth]{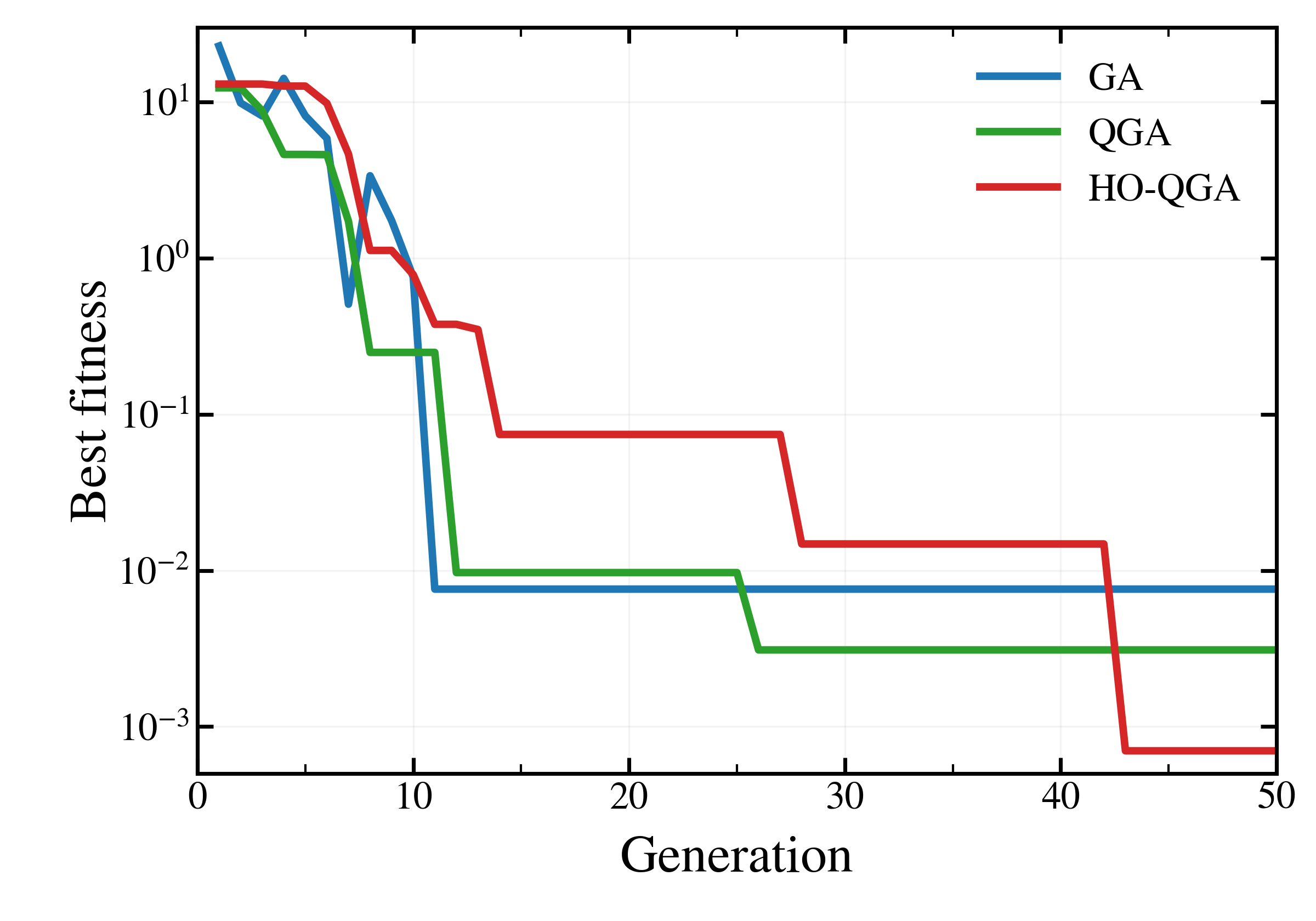}
\caption{Optimization performance of Rastrigin function using Classical genetic algorithm (GA), quantum genetic algorithm (QGA), and higher-order quantum genetic algorithm (HO-QGA)}
\label{fig:rastrigin}
\end{figure}

\noindent Fig.~\ref{fig:rastrigin} shows the results for Rastrigin function, which have a global minimum $f_{\min}=0$. Classical GA reaches $0.0076$ at generation $11$, QGA reaches $0.0031$ at generation $26$ and HO-QGA gets $0.0007$ at generation $43$. These results show that classical GA can show relatively fast improvement initially but HO-QGA is much better for final optimization accuracy.

\begin{figure}[H]
\centering
\includegraphics[width=0.73\textwidth]{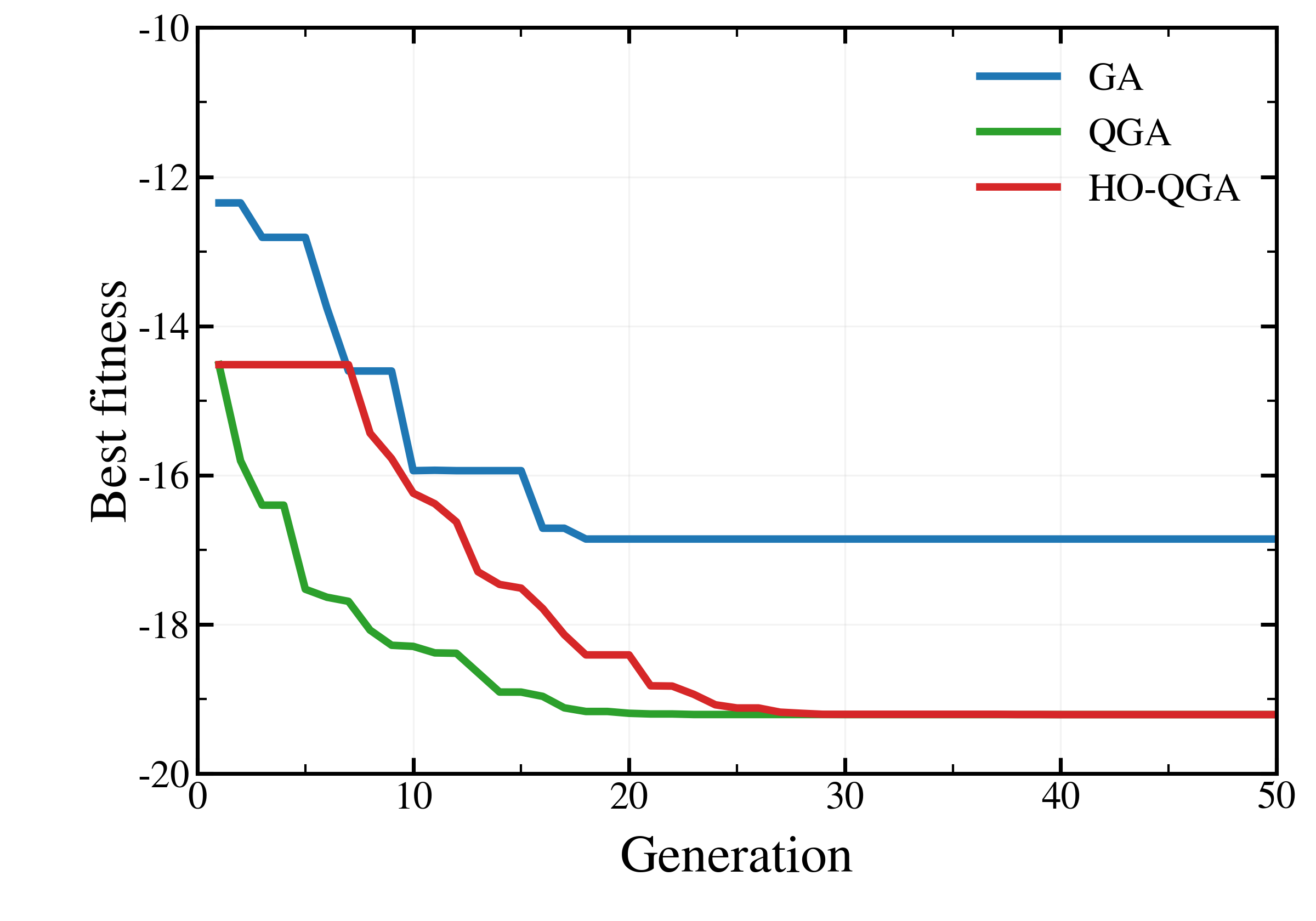}
\caption{Optimization performance of Table Holder function using Classical genetic algorithm (GA), quantum genetic algorithm (QGA), and higher-order quantum genetic algorithm (HO-QGA)}
\label{fig:table_holder}
\end{figure}

\noindent Fig.~\ref{fig:table_holder} represents the optimization of Table Holder function, which have a theoretical global minimum value $f_{\min}\simeq-19.2085$. Classical GA fitness value is $-12.3492$ initially and reaches $-16.8551$ at generation $18$, with no improvement afterwards. Comparatively, QGA gets $-19.208$ at generation $30$ and HO-QGA gets $-19.2085$ at generation $40$. Both quantum-inspired algorithms get solutions very near to global minimum with a fast convergence rate but classical GA fails in it.

Overall, the quantum-inspired approaches get more accurate final results comparative to the classical GA.

\section{Conclusions}

This research compared the performance of Classical Genetic Algorithm, Quantum Genetic Algorithm, and Higher-Order Quantum Genetic Algorithm in optimizing three benchmark functions: RosenBrock, Rastrigin and Table Holder. The experiments were performed with population size of $10$, a mutation rate of $0.1$ and a maximum of $50$ generations. The results have shown high differences in accuracy and the rate of convergence between the approaches in each benchmark function. 

The results show that classical GA worked at identifying the global minimum of the Rosen Brock function at $0.0048$ after $45$ generations while the QGA enhanced the accuracy to $0.0007$ converging much faster than classical GA and took $35$ generations to converge. As compared to other genetic algorithms, HO-QGA converged faster to a near minimal solution of $0.0002$ in $39$ generations. When using the Rastrigin function, the classical GA reached the best fitness at $0.0076$. The convergence behavior of QGA is slightly better than that of classical GA, it took $26$ generations to produce such result of $0.0031$ while HO-QGA provide better results than both and achieved the perfect minimum of $0.0007$ after $43$ generations. In case of the Table Holder function, there was a considerably more complex situation because of the high multimodality of the task. In this case, classical GA reached a local minimum of $-16.8$ while QGA provides significantly better result than classical GA reaching $-19.208$ after $30$ generations, HO-QGA excelled by matching the theoretical global minimum of $-19.2085$ after $40$ generations, demonstrating the highest accuracy and robustness.

This comparison further confirmed that HO-QGA provided the highest accurate solutions of all functions and especially in Table Holder function. It also showed faster convergence in basic terrains such as the Rosen Brock function. Compared with the other algorithms, QGA and HO-QGA showed satisfactory trade-off between precision and time consumption and also to the number of generations which must go through in order to get optimal solutions. Current study is useful for solving higher-order quantum-inspired representation complex optimization problems with vast applications in engineering optimization, machine learning, parameter estimation.


\end{document}